# Emissive Surface Traps Lead to Asymmetric Photoluminescence Line Shape in Spheroidal CsPbBr$_3$ Quantum Dots

Jessica Kline[1], Shaun Gallagher[1], Benjamin F. Hammel[2], Reshma Mathew[3], Dylan M. Ladd[2], Robert J. E. Westbrook[1], Jalen N. Pryor[4], Michael F. Toney[2,4,5], Matthew Pelton[3], Sadegh Yazdi[2,5], Gordana Dukovic[2,5,6] and David S. Ginger[1]*

[1]Department of Chemistry, University of Washington, Seattle, WA 98195, USA

[2]Materials Science and Engineering, University of Colorado Boulder, Boulder, CO 80309-0215, USA

[3]Department of Physics, University of Maryland, Baltimore County, Baltimore, Maryland 21250, USA

[4]Department of Chemical and Biological Engineering, University of Colorado Boulder, Boulder, Colorado 80303, United States

[5]Renewable and Sustainable Energy Institute, University of Colorado Boulder, Boulder, CO 80309-0215, USA

[6]Department of Chemistry, University of Colorado Boulder, Boulder, CO 80309-0215, USA

* Corresponding author: dginger@uw.edu

**Abstract:** The morphology of quantum dots plays an important role in governing their photophysics. Here, we explore the photoluminescence of spheroidal CsPbBr$_3$ quantum dots synthesized *via* the room-temperature trioctlyphosphine oxide/PbBr$_2$ method. Despite photoluminescence quantum yields nearing 100%, these spheroidal quantum dots exhibit an elongated red photoluminescence tail not observed in typical cubic quantum dots synthesized *via* hot injection. We explore this elongated red tail through structural and optical characterization including small-angle x-ray scattering, transmission electron microscopy and time-resolved, steady-state, and single quantum dot photoluminescence. From these measurements we conclude that the red tail originates from emissive surface traps. We hypothesize that these emissive surface traps are located on the (111) surfaces and show that the traps can be passivated by adding phenethyl ammonium bromide, resulting in a more symmetric line shape.

Lead halide perovskite quantum dots are a promising material for a variety of next-generation optoelectronics.[1] Lead halide perovskite quantum dots exhibit tunable emission,[2] narrow linewidths[3] and near unity photoluminescence quantum yields (PLQYs)[4] making them attractive for both quantum and classical optoelectronics. The narrow linewidths of perovskite quantum dots are particularly exciting as narrow linewidths are a prerequisite for any emission-based application– improving color purity and single-photon indistinguishability.[5] However, several factors can result in linewidth broadening and line shape asymmetry– including the chemical environment, size and morphology polydispersity, phonon interactions, fine structure, and spectral diffusion. [5,6] The absolute minimization of quantum dot linewidths requires understanding how these different factors impact the overall line shape.

At the single quantum dot limit, the passivating ligand has a strong effect on the linewidth[7–10] while in ensembles, the size dispersity has a pronounced effect on the linewidth.[6] The most effective way to control size dispersity and ligand passivation is through a high-quality synthesis that produces a narrow size distribution and includes effective surface ligand passivation.[11,12] One very promising synthesis for perovskite quantum dots that meets these criteria is the room temperature trioctylphosphine oxide (TOPO)/$PbBr_2$ method[13] which both improves quantum dot monodispersity[14] and uses facile ligand exchanges for optimal passivation.[15,16] Additionally, this synthesis produces high PLQY[13,16,17] quantum dots with faster radiative lifetimes[17] making them attractive candidates for emission-based applications. However, as previously noted,[13,17,18] $CsPbBr_3$ quantum dots from this synthesis have an asymmetric photoluminescence line shape. Herein, we explore the cause of this elongated red photoluminescence tail using structural characterization and steady-state and time-resolved photoluminescence measurements. We find that emissive traps cause the observed photoluminescence asymmetry and propose that these traps are due to the spheroidal morphology of quantum dots synthesized *via* the TOPO/$PbBr_2$ method. Finally, we demonstrate a ligand-based framework for passivating these emissive traps and increasing the line shape symmetry.

We begin by synthesizing and characterizing lecithin-capped $CsPbBr_3$ quantum dots with spheroidal[13] and cubic[19] morphologies. We synthesize four spheroidal quantum dot samples of varying sizes which are referred to as S4, S5, S7 and S12. Our cubic quantum dot sample is referred to as C9. Figure S1 and Tables S1 and S2 contain more details on the synthesis conditions for the spheroidal quantum dots. Figure S2 shows the morphologies of all quantum dots as observed with scanning transmission electron microscopy (STEM).

Figure 1a compares typical solution photoluminescence spectra from S7 (orange line) and C9 (grey line) samples which are similar in size (7.1 ± 0.6 nm diameter vs 9.3 ± 1.6 nm edge length from TEM). Figure 1a also shows Gaussian fits (dashed lines) to each spectrum and the corresponding skew values (Equation S3). The emission from C9 samples is well-described by a Gaussian, with a slight asymmetry indicated by average skew values of -0.03 ± 0.07. In contrast, we find that the S7 sample deviates more from a Gaussian line shape, exhibiting a larger tail on the low energy (red) edge of the spectrum and skew values over ten times larger (-0.50 ± 0.05) than C9. The elongated red photoluminescence tail we see in S7 quantum dots is also present in S4, S5 and S12 samples. Figure 1b is representative of the skew values we observed for all synthesized spheroidal and cubic quantum dot samples. Interestingly, we observe size-dependent line shape asymmetry with the smaller spheroidal quantum dots having larger skew values. Similarly, when we quantify the sample morphologies using ImageJ's circularity metric[20] (Figure S2), smaller quantum dots have a more spheroidal shape. This suggests that larger ensemble photoluminescence skew values are correlated to a more spheroidal morphology. Figures 1c-1f show the ensemble optical characterization for representative synthetic batches of the spheroidal quantum dots. Figure S3 shows the ensemble optical characterization of a representative batch of the C9 quantum dots.

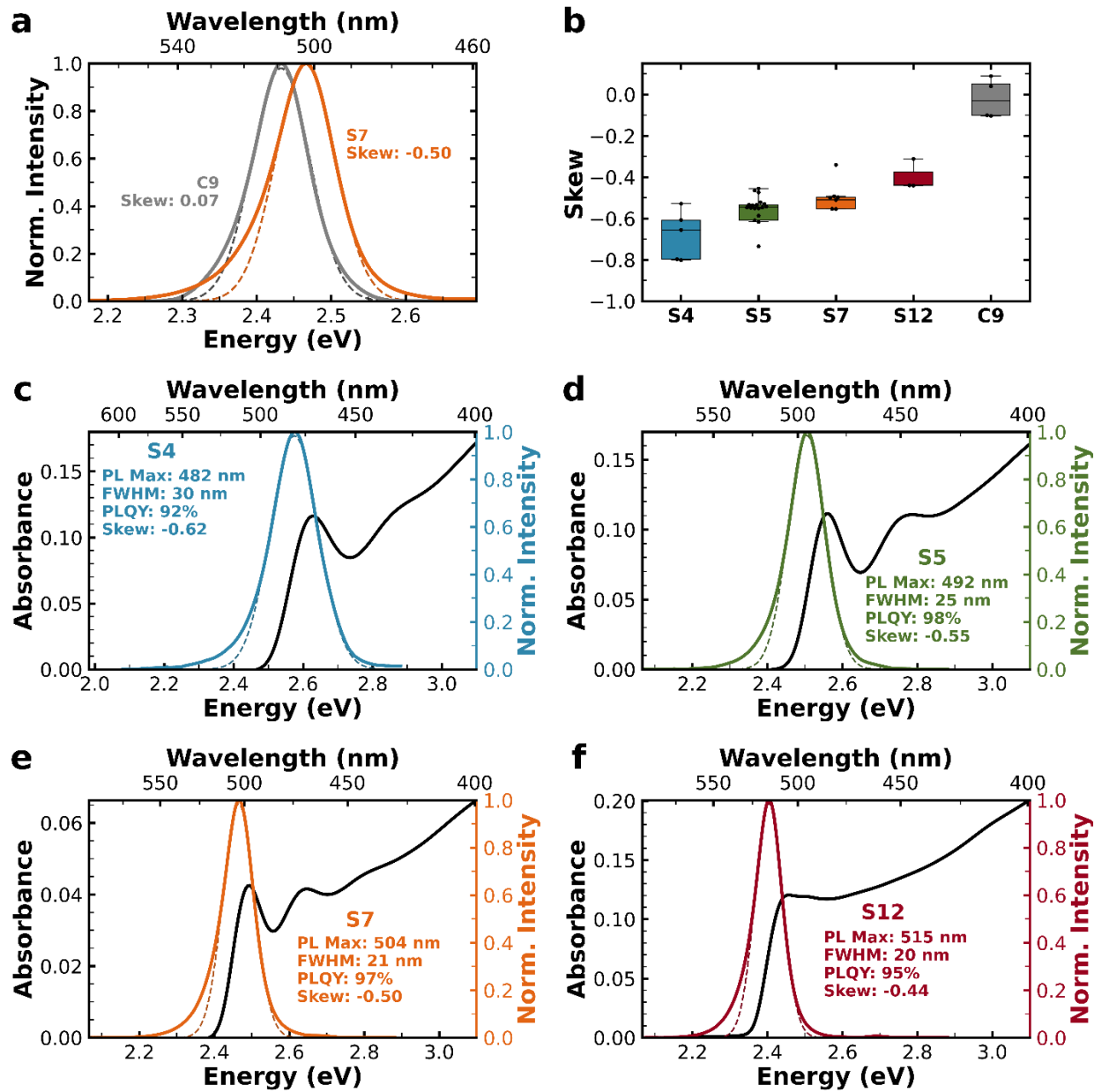

**Figure 1. Ensemble Characterization. a)** Representative photoluminescence spectra of S7 (orange line) and C9 (grey line) quantum dots and the Gaussian fits to the spectra. The S7 quantum dots are best fit by a Gaussian with a mean of 2.47 eV and standard deviation of 40 meV. The C9 quantum dots are best fit by a Gaussian with a mean of 2.43 eV and a standard deviation of 39 meV. S7 quantum dots exhibit an elongated red photoluminescence tail in contrast to the symmetric photoluminescence from C9 quantum dots. **b)** Distribution of skew values for S4, S5, S7, and S12 quantum dots compared to the skew values of C9 quantum dots. Each distribution comes from at least three synthetic batches for each type of quantum dot which are shown as black dots. Colored boxes extend from the first quartile to the third quartile, with the mean indicated by a solid line. Whiskers extend to 1.5x the interquartile range. **c)** Ensemble solution characterization for S4 quantum dots. **d)** Ensemble solution spectra for S5 quantum dots. **e)** Ensemble solution spectra for S7 quantum dots **f)** Ensemble solution characterization for S12 quantum dots. Dashed lines show Gaussian fits to the photoluminescence spectra. Absorbance has units proportional to OD·s. Photoluminescence spectra were acquired from dilute room temperature solutions excited at 405 nm. Photoluminescence has units proportional to photons·s$^3$eV$^{-1}$. We corrected all emission spectra for instrument response as described in the supporting information.

Understanding the origin of the observed line shape asymmetry is important to producing spheroidal quantum dots with the narrowest possible linewidths. One possible cause of an asymmetric line shape could be an asymmetric quantum dot size distribution.[19,21] Therefore, we measured the size distributions via small angle x-ray scattering (SAXS) and transmission electron microscopy (TEM) to look for evidence of size dispersity. Figure 2a shows the size distributions measured *via* SAXS, and Figure S4 compares the size distributions acquired by SAXS and TEM on the same samples. Figure S5 shows the fits to the SAXS scattering patterns for all spheroidal quantum dots and Table S3 contains the fit parameters.

We find that the SAXS data is well fit by a bimodal Schulz distribution which describes the primary quantum dot population and a secondary background population between 2 and 4 nm in diameter. This secondary population is indistinguishable from the background in TEM due to low contrast, however Figure S6 confirms the presence of this secondary population using STEM. We find that all considered possible identities for this secondary population are optically inactive and cannot cause the observed photoluminescence asymmetry (see supporting information for more detail).

We find that the quantum dot size distributions acquired from both methods are in good agreement, with mean diameters that are less than 9% different. Most importantly, the two sizing methods agree on the width and shape of the size distribution for all samples. While the spheroidal quantum dot populations are slightly skewed (0.22 to 0.30 according to SAXS), these skew values are only half of the photoluminescence skew values (-0.40 to -0.80) and remain approximately constant regardless of size. Additionally, according to TEM, C9 quantum dots also have a size distribution skew of 0.20 and, as shown in Figure 1b, C9 samples show significantly less ensemble photoluminescence skew than spheroidal samples. These results suggest that the asymmetric photoluminescence does not primarily result from size dispersion in the ensemble.

To further explore the potential effect of size dispersity on the ensemble skew, we also measure single quantum dot photoluminescence spectra. Figure 2b shows representative single quantum dot photoluminescence spectra from several individual particles. The spectra of the single spheroidal quantum dots display an elongated red tail (skew values of -0.65 ± 0.15), while the spectra of C9 quantum dots are more symmetric (skew values of -0.02 ± 0.08). Figure S7 shows the distribution of skew values from all single quantum dot photoluminescence spectra. We find that the average skew of a single quantum dot spectrum is within 0.05 of the ensemble skew for all samples, confirming that the ensemble skew is primarily due to the photophysical properties of the individual quantum dot and not size dispersion.

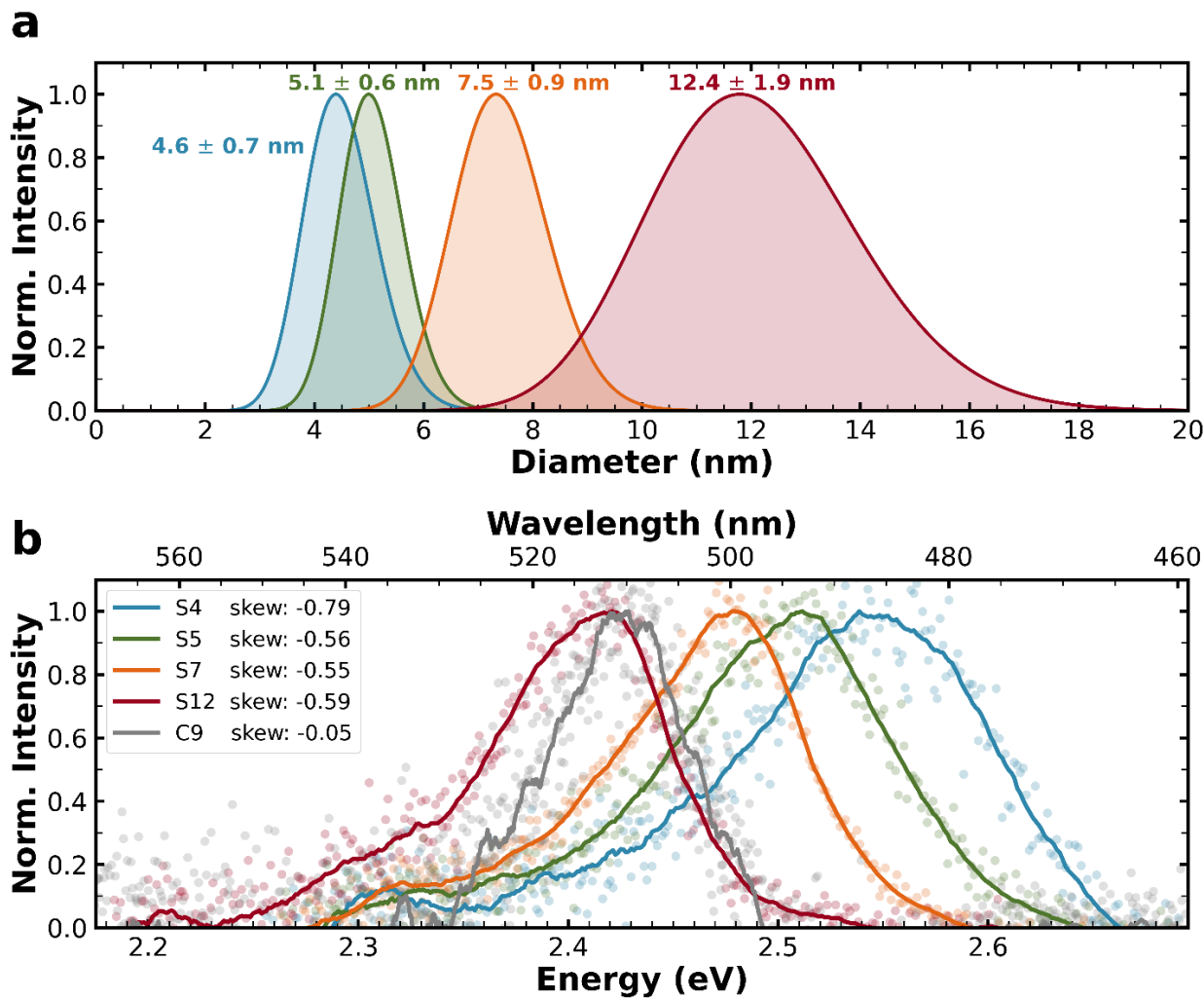

**Figure 2. Photoluminescence asymmetry is not an ensemble effect. a)** SAXS size distributions for spheroidal quantum dots. Size distributions are best described by a bimodal Schulz distribution which represents the larger quantum dot population along with a population of smaller particles at 2-4 nm. The secondary population is not shown here (see Table S3 and Figure S6 for details on this population). The average quantum dot diameter and standard deviation are written above the distributions. **b)** Single quantum dot photoluminescence spectra for spheroidal and cubic quantum dots. Dots represent the raw data and solid lines the smoothed data. Spheroidal quantum dots exhibit elongated red tails and have skew values between -0.80 and -0.50. C9 quantum dot spectra are more symmetric with skew values between +0.08 and -0.10. Quantum dots were drop-cast onto glass coverslips with no polymer matrix, and photoluminescence was measured at room temperature under nitrogen using 420 nm excitation. Emission maxima and line widths match well with ensemble characterization (see Table S4 for more details).

After eliminating photoluminescence asymmetry arising from the size distribution as the primary cause of the elongated red tail, we consider other potential causes. First, we consider whether a strong biexciton contribution to the ensemble photoluminescence causes the elongated red tail. Such an effect could lead to a red tail as biexciton emission is a lower energy process than excitonic emission.[22–24] In that case, we expect the photoluminescence asymmetry to increase with higher excitation powers. To look for evidence of biexciton emission, we acquired photoluminescence spectra with varying excitation powers. Figure S8 shows the photoluminescence of the S7 quantum dots measured at powers from 0.95 mW/cm$^2$ to 170 mW/cm$^2$. The photoluminescence intensity varies linearly with excitation power over four orders of magnitude with no observable change in line shape. We thus rule out biexciton emission as the cause of the observed photoluminescence asymmetry under our experimental conditions.

Next, we consider that the change in quantum dot morphology results in different exciton fine structure, introducing a second emissive state.[25–27] In that case, we expect to see differences in the absorption fine structure of spheroidal and cubic quantum dots. We use photoluminescence excitation spectroscopy to monitor and compare band-edge fine structure. Figure S9 compares photoluminescence excitation spectra from S7, S12 and C9 samples. The photoluminescence excitation spectra show no difference in band-edge fine structure between these cubic and spheroidal quantum dots. We thus rule out exciton fine structure as the cause of the observed photoluminescence asymmetry in spheroidal quantum dots.

Having ruled out a secondary excitonic contribution to the photoluminescence, we now consider band-edge disorder related causes, including emissive traps[22,23,28] and exciton-phonon coupling.[29] In a disordered system, the band edge can vary spatially or temporally, resulting in potential minima, which form a band-tail of shallow states associated with phonon modes and traps.[28,30]

If band-edge disorder is the cause of the elongated photoluminescence tail, we expect to see signs of disorder in the absorbance spectrum. Specifically, the onset of the absorption band-edge contains information about the density of band-tail states. We quantify the slope of the absorption onset in our quantum dot samples using an Urbach tail fit. Shallower slopes indicate increased disorder and a higher density of band-tail states. As the absorbance spectrum of a quantum dot sample contains contributions from excitonic and continuum components, the Urbach tail cannot be determined simply by fitting the measured absorbance onset. Thus, we use an Elliot model function to separate excitonic and continuum contributions.[31] Figure S10 shows Elliot model fits to the quantum dot absorbance spectra in Figure 3a, and Table S5 contains the fit parameters. Figure 3a shows the measured absorbance (solid lines) and the continuum absorbance contributions (dashed lines) for both spheroidal and cubic quantum dots. We characterized the Urbach tails by fitting the onset of the continuum contribution to Equation 1.[31]

$$A_{continuum}(E) = A_0 \sqrt{\frac{k_B T}{2 m_{urbach}}} \exp\left[\frac{m_{urbach}}{k_B T}(E - E_g) - \frac{1}{2}\right] \quad (1)$$

$A_0$ is a prefactor with units of nm·s·eV$^{-3/2}$, $k_B$ is the Boltzmann constant, T is the temperature, $m_{urbach}$ describes the slope of the exponential onset and $E_g$ is the optical bandgap. Figure S11 shows the fits of Equation 1 to the continuum absorbance contributions. Within the spheroidal quantum dot samples, the onset slope exhibits a clear trend, with smaller quantum dots having a shallower onset. The S4 sample has the shallowest onset ($m_{urbach}$ = 0.899 ± 0.005), and the S12 sample has the steepest onset ($m_{urbach}$ = 0.942 ± 0.003) indicating that band-edge disorder scales with size. This is an expected trend, as the strongest contributing factor to band-edge disorder in perovskites is exciton-phonon coupling[29] which increases in strength as quantum dot size decreases.[32] However, the C9 quantum dots do not fit the size-based trend seen in the spheroidal quantum dots; despite the mid-range size, the C9 sample has a steeper onset ($m_{urbach}$ = 0.994 ± 0.004) than either the S7 or S12 quantum dots. The deviation of the C9 sample from the trend in spheroidal quantum dots indicates that spheroidal quantum dots have significantly more band-edge disorder than cubic quantum dots. We consider two likely causes for the increased band-edge disorder in spheroidal quantum dots: increased exciton-phonon coupling strength and increased trap density.

We turn to emission-energy-dependent photoluminescence lifetimes to distinguish between band-tail states resulting from exciton-phonon coupling and trap density. If the elongated red tail is caused by increased exciton-phonon coupling strength, we expect to see a lifetime independent of emission energy.[29] However, if the elongated red tail is caused by radiative recombination at trap states, the lifetime should increase as the emission energy decreases and the traps become deeper.[22,23,33] Figure 3b shows emission-energy-dependent lifetimes for the spheroidal quantum dots acquired *via* streak camera (see supporting information section on streak camera measurements for details). Lifetimes were fit by a stretch exponential (Equation S4) to capture the distribution of recombination dynamics present in perovskites.[34] Figure S12 shows the photoluminescence lifetime data and fits for selected emission energies. In the spheroidal quantum dot samples, the lifetime increases as the emission energy decreases, which suggests a red-edge emission that arises from emissive traps.

To directly probe for emissive traps, we collected excitation-energy-dependent photoluminescence spectra. We compare spectra collected with above-gap excitation, which creates a population of excitons that primarily recombine at the band-edge, and sub-gap excitation, which creates an initial population of excitons in the band-tail states, to distinguish between trap-mediated emission and band-edge emission. If the spectrum acquired with sub-gap excitation is identical to its above-gap counterpart, then the initial population of band-tail excitons up-convert and recombine at the band-edge.[35,36] If the spectrum acquired with sub-gap excitation is red-shifted and broader than its above-gap counterpart, then radiative recombination occurs in the band-tail states.[37] Figure 3c shows selected photoluminescence spectra of the quantum dot samples acquired using above-gap and sub-gap excitation. Table S6 compares the emission maxima and line widths from above-gap and sub-gap excitation, and Figure S13 shows additional excitation energy-dependent spectra from the quantum dot samples. The spectra from C9 quantum dots with sub-gap and above-gap excitation are identical – indicating there is no emission from trap states in cubic quantum dots. In contrast, the spectra from spheroidal quantum dots with sub-gap excitation are red-shifted and broader than their above-gap counterparts – indicating that spheroidal quantum dots have emissive traps.

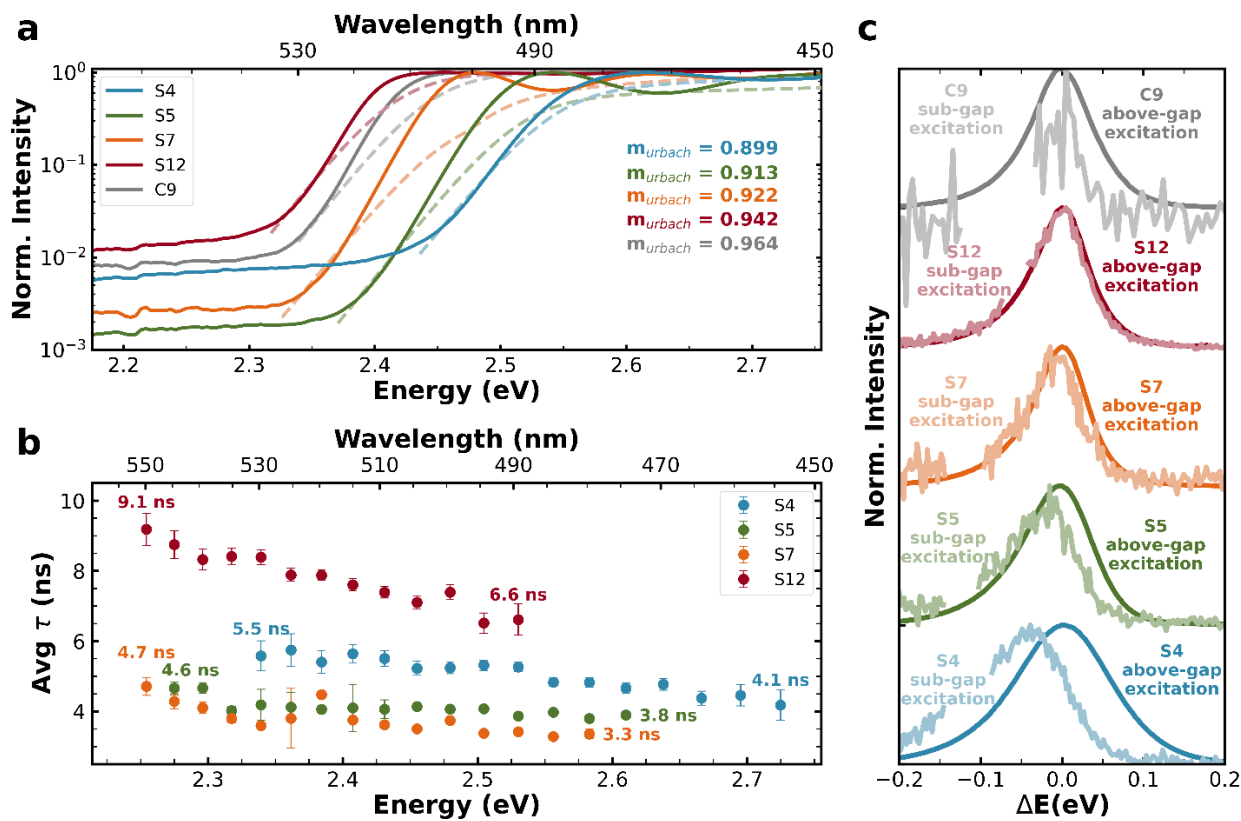

**Figure 3. Evidence for Traps. a)** Absorbance spectra for spheroidal and cubic quantum dots. The contributions of the continuum absorbance are shown with a dashed line. The slope of the continuum onset ($m_{urbach}$) becomes shallower with decreasing quantum dot size for spheroidal quantum dots and is the steepest in cubic quantum dots. **b)** Emission energy dependent photoluminescence lifetimes for spheroidal quantum dots. The lifetimes increase as the emission energy decreases indicating emission from traps on the low energy side. **c)** Overlayed photoluminescence spectra acquired with above-gap or sub-gap excitation for spheroidal and cubic quantum dots. Sub-gap excitation photoluminescence in spheroidal quantum dots arises from emissive traps. Spectra labeled above-gap excitation were acquired by exciting at 400 nm. Spectra labeled sub-gap excitation were acquired by exciting at 510 nm, 520 nm, 530

nm, 540 nm, and 530 nm, respectively, for S4, S5, S7, S12, and C9 quantum dots. Scatter from the sub-gap excitation was removed from the spectrum for clarity.

If surface traps cause the observed photoluminescence asymmetry, passivating these traps should reduce the asymmetry. However, as Figure S14 shows, none of our selected common ligand systems reduce the line shape asymmetry, and we must further consider the surface of the quantum dots and our choice of ligand.

In considering the origin of emissive traps in spheroidal quantum dots, we suggest that the faceted nature of spheroidal quantum dots deserves more consideration. Cubic quantum dots contain only (100) facets while spheroidal quantum dots have (100), (110) and (111) facets.[13,17] Figure S15 shows illustrations of each of these facets, which provide unique surfaces to be passivated by the ligands. Most importantly, the polarity of these facets differ significantly from each other.[38–42] In particular, passivation of the highly polar (111) facet is strongly correlated to a ligand's polarizability.[38] Lecithin is a weakly polarizable ligand, suggesting spheroidal quantum dots may have poorly passivated polar facets. Hypothesizing that the observed emissive traps are located on poorly passivated polar facets, we treat the quantum dots with phenethyl ammonium bromide (PEABr), a highly polarizable ligand that has been used for perovskite passivation. Figure S16 shows the chemical structures of lecithin and PEABr.

Figure 4a shows the photoluminescence of S7 and S12 samples passivated by lecithin and lecithin/PEABr. Figure S17 shows the photoluminescence and absorbance spectra of S4 and S5 quantum dots passivated with lecithin and lecithin/PEABr, and Figure S18 shows TEM images of lecithin-capped quantum dots before and after PEABr addition. Treating the spheroidal quantum dots with PEABr reduces the photoluminescence skew and increases the PLQY. Additionally, emission from lecithin/PEABr-capped quantum dots is blue-shifted by 6 meV, consistent with the 0.5 nm decrease in diameter observed *via* TEM (Figure S18). These changes in photoluminescence are consistent with increased passivation for lecithin/PEABr-capped quantum dots, however the increased passivation may not be unique to PEABr. Figure S19 compares lecithin-capped quantum dots treated with PEABr, $ZnBr_2$, oleylammonium, and lecithin. We find that only PEABr treatment results in an observable change to the photoluminescence – indicating the importance of the phenethyl ammonium moiety in increasing passivation.

Figure 4b shows the absorbance spectra of S7 and S12 lecithin- and lecithin/PEABr-capped quantum dots. Lecithin/PEABr-capped quantum dots have slightly steeper absorption onsets than their lecithin-capped counterparts, consistent with increased passivation from adding PEABr. Additionally, the absorbance spectrum of S12 lecithin/PEABr-capped quantum dots is slightly blue-shifted from their lecithin-capped counterparts, consistent with the change in diameter observed in TEM (Figure S18). We also observe increased absorbance for S7 lecithin/PEABr-capped quantum dots consistent with the more cubic morphology seen in TEM (Figure S18). For equal edge-lengths and diameters, a cube has approximately twice the volume of a sphere, and as such the morphological volume change affects the absorption spectrum more than the decreased diameter. We acknowledge that this morphology change likely plays a role in the decreased photoluminescence asymmetry for S7 lecithin/PEABr-capped quantum dots, however as S12 lecithin/PEABr-capped quantum dots show no observable morphology change and decreased photoluminescence asymmetry, we suggest that PEABr still plays an important role in increasing passivation.

Figures 4c and 4d compare photoluminescence lifetimes for the two ligand systems in S7 and S12 quantum dots. Lecithin/PEABr-capped quantum dots have a shorter lifetime and larger stretching exponent than their lecithin-capped counterparts. Table S7 shows that the observed lifetime reduction is primarily due to increased radiative recombination rates. Additionally, the increased stretching exponent indicates that the lecithin/PEABr-capped sample has a more homogenous distribution of lifetimes.[34] Altogether, we find that adding PEABr reduces the elongated red photoluminescence tail and electronic disorder and increases radiative recombination rates. These trends in photoluminescence line shape, absorbance, and lifetime present a self-consistent story in which PEABr provides additional passivation to spheroidal quantum dots.

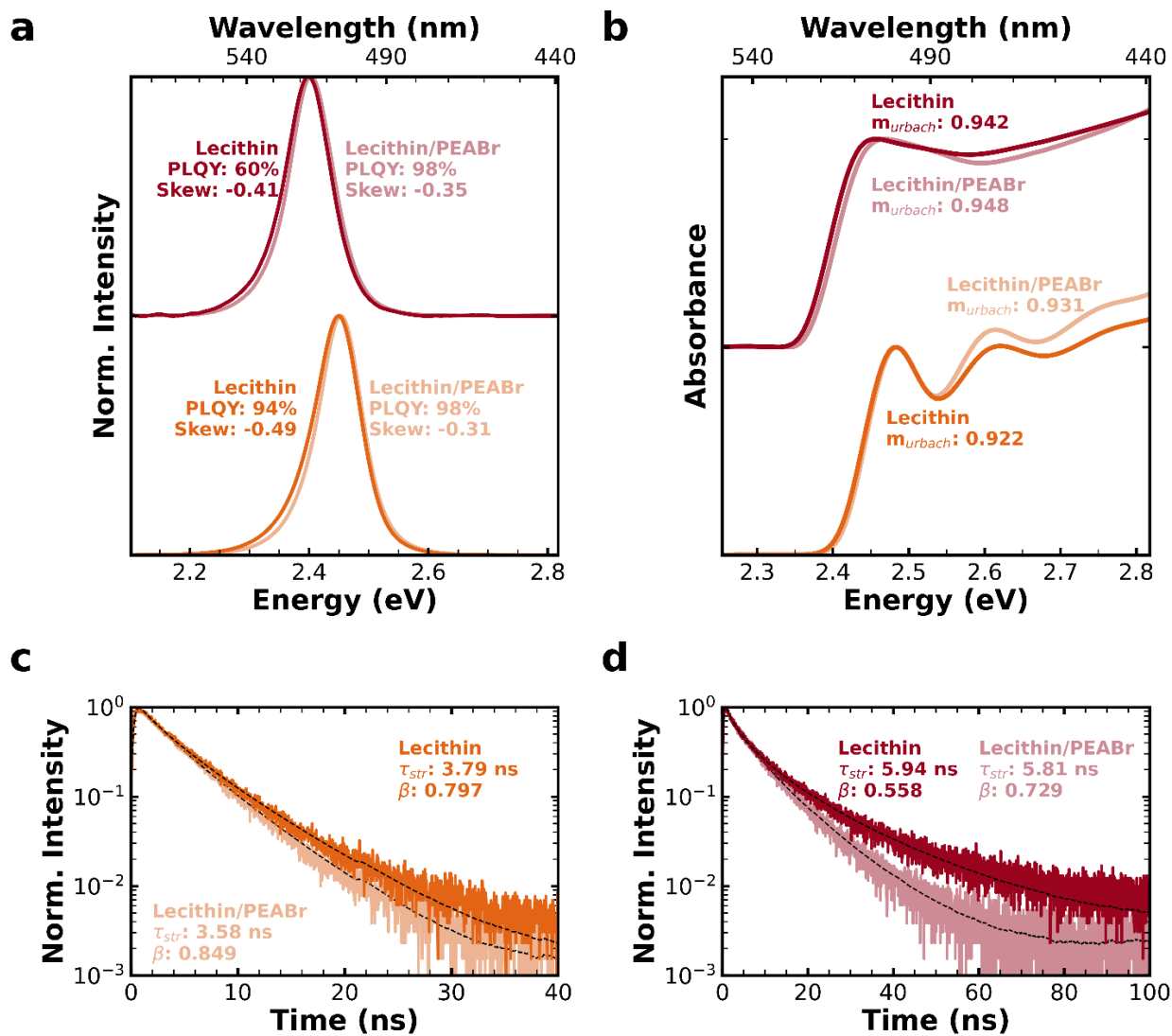

**Figure 4. Passivating Emissive Traps on the Spheroidal Quantum Dot Surface. a)** Photoluminescence spectra for S12 (red) and S7 (orange) lecithin- and lecithin/PEABr- capped spheroidal quantum dots. Capping with lecithin/PEABr results in increasingly symmetric spectra with higher PLQYs, indicating increased passivation. S7 lecithin-capped quantum dot emission is centered at 2.445 ± 0.044 eV, and lecithin/PEABr-capped quantum dot emission is centered at 2.451 ± 0.041 eV. S12 lecithin-capped quantum dot emission is centered at 2.397 ± 0.043 eV and lecithin/PEABr-capped quantum dot emission is centered at 2.403 ± 0.041 eV. The decreased PLQY for lecithin-capped S12 quantum dots relative to those reported in Figure 1f (95% vs 60%) is caused by ligand stripping from additional wash steps before adding PEABr.[43] **b)** Absorbance spectra for S12 (red) and S7 (orange) lecithin- and lecithin/PEABr- capped spheroidal quantum dots. The Urbach slope is slightly steeper for lecithin/PEABr-capped quantum dots. Photoluminescence lifetimes for **c)** S7 and **d)** S12 lecithin- and lecithin/PEABr-capped spheroidal quantum dots. Lecithin/PEABr-capped quantum dots see reduced lifetimes and increased stretching exponents, indicating the passivation of the emissive traps.

While adding PEABr passivates the emissive defects present in lecithin-capped spheroidal quantum dots, PEABr is not an ideal ligand for long-term quantum dot passivation. Figure S20 shows that PEABr addition results in the formation of quasi-two-dimensional perovskite byproducts, and Figure S21 shows that PEABr decreases storage stability. However, the reduction of the elongated red tail after PEABr addition indicates the promise of similar ligands and the importance of facet-specific investigations of ligand binding in perovskite quantum dots.

Through a variety of steady-state and time-resolved spectroscopies, we study the asymmetric red photoluminescence tail of spheroidal $CsPbBr_3$ quantum dots. Single quantum dot photoluminescence spectra, SAXS, and TEM confirm that an asymmetric size distribution does not cause the observed asymmetric photoluminescence tail. Based on emission energy dependent lifetimes, Urbach tail fittings, and photoluminescence spectra with sub-gap excitation, we conclude that emissive traps are responsible for the observed photoluminescence asymmetry. Upon considering the different facets of spheroidal quantum dots, we suggest that common ligands may struggle to passivate at least one of the different facets and consider a mixed ligand system of lecithin and PEABr to passivate spheroidal quantum dots better. Lecithin/PEABr-capped quantum dots show decreased photoluminescence asymmetry and increased lifetime homogeneity. We also note that PEABr is not an ideal ligand for perovskite quantum dot passivation due to the formation of by-products and storage instability. These findings indicate the promise of other highly polarizable ligands, such as N,N-diphenacyl oleylammonium, aromatic phosphoethanolamine derivatives and 1-(*p*-tolyl)ethylamine, for passivating spheroidal quantum dots. They also highlight the drastic impact of morphology and faceting on the fundamental optical properties of perovskite quantum dots.

**Supporting Information**

Additional experimental details, synthesis methods, and characterization.

**Author Contributions**

The manuscript was written through contributions of all authors. All authors have given approval to the final version of the manuscript.

**Acknowledgements**

This work, and the roles of J.K., S.G., B.F.H, R.M., D.M.L., J.N.P., M.F.T., M.P., S.Y., G.D. and D.S.G were primarily supported by the National Science Foundation under the STC IMOD Grant (No. DMR-2019444). B.F.H. and D.M.L. acknowledge support from the National Science Foundation through the Graduate Research Fellowship Program (NSF-GRFP) under Grant No. DGE 2040434. R.J.E.W. carried out streak camera measurements and was supported by the Office of Naval Research (ONR N000-14-20-1-2191) and the Momental Foundation via the Mistletoe Fellowship. The authors acknowledge the use of facilities and instruments at the Photonics Research Center (PRC) at the Department of Chemistry, University of Washington, as well as that at the Research Training Testbed (RTT), part of the Washington Clean Energy Testbeds system. Part of this work was carried out at the Molecular Analysis Facility, a National Nanotechnology Coordinated Infrastructure site at the University of Washington which is supported in part by the National Science Foundation (NNCI-1542101), the Molecular Engineering & Sciences Institute, and the Clean Energy Institute. TEM was carried out at the Facility for Electron Microscopy of Materials at the University of Colorado Boulder (CU FEMM, RRID: SCR_019306).

J.K. acknowledges David M. Jonas (professor, University of Colorado Boulder) for discussion regarding potential causes of an elongated red photoluminescence tail in quantum dots.

D.M.L. acknowledges use of the SasView application for fitting SAXS data. SasView was originally developed under NSF award DMR-0520547 and contains code developed with funding from the European Union's Horizon 2020 research and innovation program under the SINE2020 project, grant agreement No 654000.

B.F.H. acknowledges Olivia F. Bird (graduate student, University of Colorado Boulder) and Sophia M. Click (postdoctoral researcher, University of Colorado Boulder) for discussions related to TEM image segmentation and size analysis using Trainable Weka Segmentation in ImageJ.

The authors declare no competing financial interests.